\documentclass[epj]{webofc}
\usepackage[utf8]{inputenc}
\usepackage[varg]{txfonts}   
\usepackage{booktabs}
\usepackage{xcolor}
\definecolor{darkred}{rgb}{0.4,0.0,0.0}
\definecolor{darkgreen}{rgb}{0.0,0.4,0.0}
\definecolor{darkblue}{rgb}{0.0,0.0,0.4}
\usepackage[bookmarks,linktocpage,colorlinks,
    linkcolor = darkred,
    urlcolor  = darkblue,
    citecolor = darkgreen]{hyperref}
\wocname{EPJ Web of Conferences}
\woctitle{Lattice2017}
\usepackage[T1]{fontenc}
\usepackage{subdepth}
\usepackage{microtype}
\usepackage{mciteplus}
\usepackage[toc,page]{appendix}
\usepackage{nicefrac}
\usepackage{subcaption}

\newcommand{\ii}{{\boldsymbol{\hat{\imath}}}}
\newcommand{\jj}{{\boldsymbol{\hat{\jmath}}}}
\newcommand{\xx}{{\mathbf{x}}}

\newcommand{\ud}{\mathrm{d}}

\newcommand{\ah}{\mathrm{ah}}

\newcommand*{\rom}[1]{\expandafter\@slowromancap\romannumeral #1@}
\begin{document}
\title{%
Plasmon mass scale and quantum fluctuations of classical fields on a real time lattice
}
\author{%
\firstname{Aleksi} \lastname{Kurkela}\inst{1,2}\fnsep \and
\firstname{Tuomas} \lastname{Lappi}\inst{3,4} \and
\firstname{Jarkko}  \lastname{Peuron}\inst{3}\fnsep\thanks{Speaker, \email{jarkko.t.peuron@student.jyu.fi}}
}
\institute{%
Theoretical Physics Department, CERN, Geneva, Switzerland
\and
Faculty of Science and Technology, University of Stavanger, 4036 Stavanger, Norway
\and
Department of Physics, P.O. Box 35, 40014 University of Jyväskylä, Finland
\and
Helsinki Institute of Physics, P.O. Box 64, 00014 University of Helsinki, Finland
}
\abstract{%
  Classical real-time lattice simulations play an important role in understanding non-equilibrium phenomena in gauge theories and are used in particular to model the prethermal evolution of heavy- ion collisions. Above the Debye scale the classical Yang-Mills (CYM) theory can be matched smoothly to kinetic theory. First we study the limits of the quasiparticle picture of the CYM fields by determining the plasmon mass of the system using 3 different methods. Then we argue that one needs a numerical calculation of a system of classical gauge fields and small linearized fluctuations, which correspond to quantum fluctuations, in a way that keeps the separation between the two manifest. We demonstrate and test an implementation of an algorithm with the linearized fluctuation showing that the linearization indeed works and that the Gauss's law is conserved.
}
\maketitle

\section{Introduction}
The Color Glass Condensate (CGC) \cite{Iancu:2003xm} is an effective theory of QCD in the high energy (weak coupling) limit. In particular CGC predicts that gluon states of nonperturbatively high occupation numbers (> $\nicefrac{1}{g^2}$) are created in the initial stages of ultrarelativistic heavy ion collisions \cite{Lappi:2003bi,Lappi:2006fp,Gelis:2010nm}.  The highly occupied gluon states can be mathematically described using classical fields. The time-evolution of the classical system is given by the CYM equations
\begin{equation}
\left[D_\mu , F^{\mu \nu} \right] = 0. 
\end{equation}
We can neglect the fermion degrees of freedom because of the Pauli principle - the occupation number of the quark states can never exceed unity. 

Due to the expansion of the system, the typical occupation numbers of the gluon states decrease over time. When they become perturbative, the system admits a kinetic theory \cite{Arnold:2002zm} description. The kinetic theory evolution is followed by hydrodynamical evolution. Hydrodynamics, however, requires that the system is in local thermal equilibrium. Thus one of the most difficult problems in the field of ultrarelativistic heavy ion collisions is understanding the fast thermalization of the system \cite{Berges:2012ks}. Successful hydrodynamical description of the quark gluon plasma indicates, that thermalization happens in a time of less  than $1\nicefrac{\mathrm{fm}}{\mathrm{c}}$ \cite{Aamodt:2010pa,Huovinen:2003fa,Kolb:2000fha,Hirano:2010jg}. Recent studies have shown, that kinetic theory can be matched smoothly to hydrodynamics, and the hydrodynamical regime can be reached within very short timescales using kinetic theory \cite{Kurkela:2015qoa}. After subsequent hydrodynamical evolution the system expands and cools down, and finally freezes out. 

In ''standard'' lattice QCD (LQCD) one rotates to imaginary time and then tries to evaluate the path integral of the system. In this way one can compute equilibrium observables. We are interested in the time-evolution of the system in real time, and thus this standard approach is not applicable. However using classical approximation (which is valid in the highly occupied regime) one can do the time-evolution in a very straightforward manner. One has to simply take the classical lattice Lagrangian, and derive the classical equations of motion:
\begin{eqnarray}
\label{eq:Adot}
\dot{U}_i(\xx) &=& i E^i(\xx) U_i(\xx) 
\\
a^2 \dot{E}^i(\xx) &=& - \sum_{j\neq i } \left[\Box_{i,j}(\xx) + \Box_{i,-j}(\xx) \right]_\ah. \label{eq:Edot}
\end{eqnarray}
Here the plaquette variables are defined as 
\begin{align}
 \Box_{i,j}(\xx) &= U_i(\xx) U_j(\xx+\ii) U^\dag_i(\xx+\jj)U^\dag_j(\xx) \\
 \Box_{i,-j}(\xx) &=  U_i(\xx) U^\dag_j(\xx+\ii-\jj) U^\dag_i(\xx-\jj)U_j(\xx-\jj)
\end{align}

In numerical computations we use the SU(2) symmetry group. This simplifies the numerical work, and the results should be qualitatively similar to SU(3) \cite{Berges:2017igc,Ipp:2010uy,Berges:2008zt,Krasnitz:2001qu}. 

\section{Plasmon mass}
From the kinetic theory point of view the interesting question is, whether we can understand these color fields in a quasiparticle picture  \cite{Lappi:2016ato}. We identify the field modes of the classical fields as plasmons. Based on a kinetic theory analysis \cite{Kurkela:2012hp} one expects the plasmon mass scale to evolve in time according to $t^{\nicefrac{-2}{7}}$ power law  when starting from an initial condition with a strongly overoccupied and UV finite isotropic distribution of gluons. In order to measure the plasmon mass we use three different methods. 
\subsection{Effective dispersion relation (DR)}
The first method is to use an effective dispersion relation \cite{Krasnitz:2000gz,Berges:2012ev,Lappi:2003bi}, defined as 
\begin{equation}
\omega^2\left(k \right) = \dfrac{\left<\left|\dot{E}\left(k\right)\right|^2\right>}{\left<\left|E\left(k\right)\right|^2\right>}.
\end{equation}

In practice we extract the plasmon mass by doing a linear fit to the dispersion relation and extrapolating to zero momentum. 

\subsection{Uniform electric field method (UE)}
The second method is a measurement which probes the response of the system to an external perturbation. We add a spatially uniform chromoelectric field \cite{Kurkela:2012hp} (corresponding to a perturbation with zero momentum) on top of the original field, and then we measure the oscillation frequency between electric and magnetic energies.

\subsection{Perturbation theory, Hard Thermal Loop (HTL)} 
The third method is to use perturbation theory. In the Hard Thermal Loop (HTL) formalism the plasmon mass is given by 
\begin{equation}
\omega_{\mathrm{pl}}^2 = \frac{4}{3} g^2 N_c \int \frac{\mathrm{d}^3k}{\left(2 \pi \right)^3} \frac{f\left(k\right)}{\left|k\right|},
\end{equation}
where $f(k)$ is the quasiparticle distribution extracted from our simulation. The agreement between these methods is not clear a priori. The HTL formula assumes that the hard modes in the classical simulation behave like charged particles and give the dominant contribution to the plasmon mass. In this sense we are also testing this assumption here (However, the HTL method is probably the most commonly used way to measure plasmon mass in CYM thermalization studies \cite{Kurkela:2012hp,Berges:2017igc,Berges:2013fga}). 

\subsection{Results}
\begin{figure}
\centering
\includegraphics[scale = 0.9]{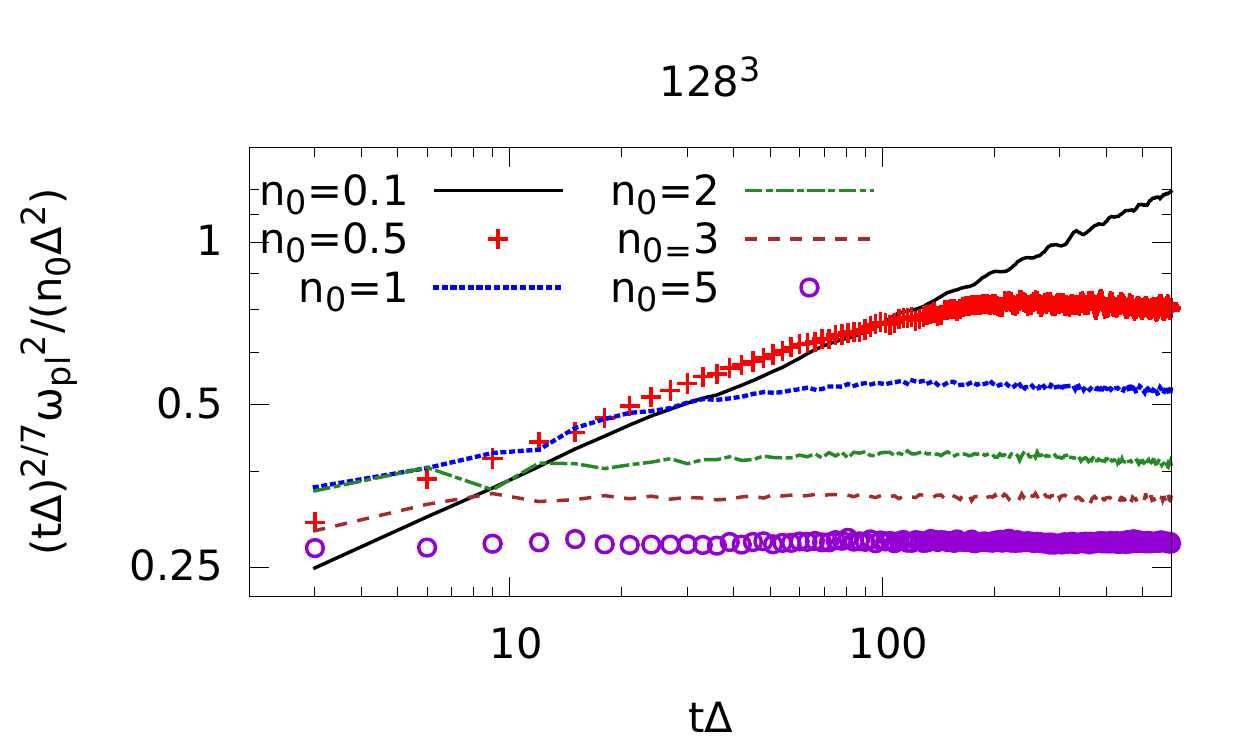}
\caption{Time dependence of the plasmon mass obtained form our simulation, scaled by the inverse of the proposed $t^{\nicefrac{-2}{7}}$ power law. We find that the late time behavior is compatible with this power law and that simulations with larger occupation numbers settle faster to this asymptotic behavior.}
\label{fig:HTLtevol}
\end{figure}

\begin{figure}
\centering
\includegraphics[scale = 0.9]{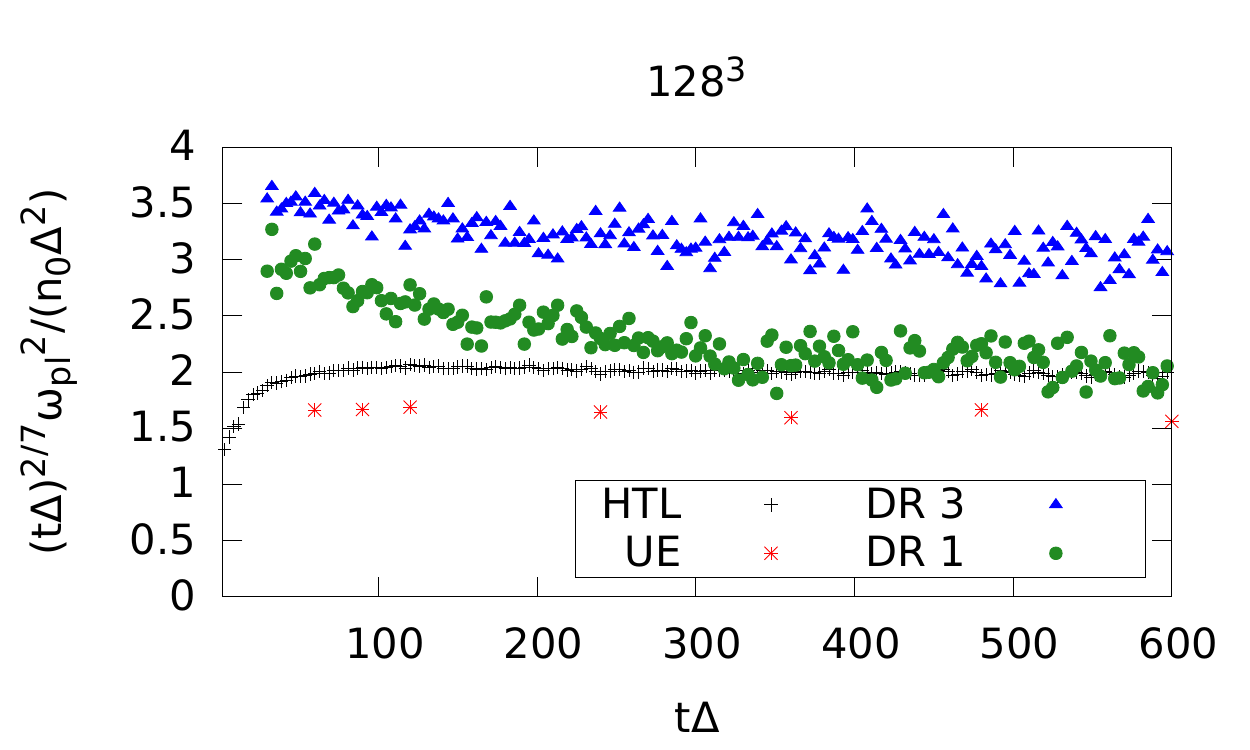}
\caption{Comparison of all three methods. Here DR 3 and DR 1 refer to the upper limits of the in the dimensionless momentum squared ($\nicefrac{k^2}{\Delta^2}$) used in the linear fit done to the dispersion relation.}
\label{fig:tevol}
\end{figure}

The results we obtain are shown in figures \ref{fig:HTLtevol} and \ref{fig:tevol}. Figure \ref{fig:HTLtevol} shows the time-evolution of the plasmon mass scale using the HTL method. The parameter $n_0$ gives the typical occupation number used in the initial condition, and $\Delta$ is the characteristic momentum scale set by the initial condition. The scale $\Delta$ should be thought of as analogous to the saturation scale $Q_s$ \cite{Gyulassy:2004zy}. We observe that the more dense systems exhibit the $t^{\nicefrac{-2}{7}}$ behavior faster than the more dilute ones. Figure \ref{fig:tevol} shows the time-evolution of the plasmon mass scale using all three methods. We find that all methods agree on the proposed power law at late times. However, the dispersion relation is dependent on the maximum momentum used in the fit. To illustrate this effect we show 2 different curves with different cutoffs. The main result in here is that the HTL and uniform electric field methods can be brought into rough agreement, but the dispersion relation method is problematic.

\section{Linearized fluctuations in CYM}
Over last couple of years, the impact of fluctuations on the space-time evolution of an ultrarelativistic heavy ion collision has been studied. It has been shown that rapidly growing quantum fluctuations \cite{Fukushima:2006ax,Dusling:2010rm,Epelbaum:2011pc,
Dusling:2012iga,Epelbaum:2013waa}  may have a remarkable impact on the isotropization of a classical Yang-Mills system \cite{Gelis:2013rba}. In these computations the fluctuations have been absorbed into the the classical background field. However, this kind of treatment of fluctuations is problematic because of two reasons. First one is that including fluctuations into the classical background is really justified only for the modes which grow large and become classical. The second problem is that one can not take continuum limit due to the fact that the fluctuation spectrum is highly UV-divergent. This means that on a fine enough lattice the UV tail of the spectrum entirely dominates the system.

Our aim is to improve the numerical treatment of these fluctuations by explicitly linearizing the fluctuation on top of the classical background and thus eliminating the backreaction from the fluctuations to the classical background \cite{Kurkela:2016mhu}. This approach has two main applications in the future. The first one is simulating the boost invariance breaking quantum fluctuations and their contribution to thermalization. The second one is to use fluctuations to study linear response in classical Yang-Mills theory. For example one can study  the dispersion relation and the damping rate of the plasma using fluctuations.

\subsection{Equations of motion for fluctuations}
Just like the background field, our fluctuation is split into two components: the fluctuation of the electric field and the fluctuation of the gauge field. Fluctuations will also have their own Gauss' law. 

\subsubsection{Fluctuation of the electric field}
We start by defining the required gauge transformation properties as those of an adjoint representation scalar field:
\begin{eqnarray}
 a_i(\xx) &\to& V(\xx) a_i(\xx) V^\dag(\xx)
\\
 e^i(\xx) &\to& V(\xx) e^i(\xx) V^\dag(\xx).
\end{eqnarray}
From these it follows that $a_i$ has to correspond to a variation of the link  matrix $U_i(\xx)$ on the left:
\begin{equation}
 U_i(\xx)_\textup{bkg + fluct} = e^{i a_i(\xx)}U_i(\xx) \approx
U_i(\xx) + i a_i(\xx)U_i(\xx).
\end{equation}
The electric field fluctuation is introduced by adding a linear perturbation to the classical background. One simply replaces $E^i$ with $E^i + e^i$. We choose to discretize the perturbation of the electric field by linearizing equation (\ref{eq:Edot}).
\begin{align} \label{eq:estep}
a^2 e^i(t+\ud t) & = a^2 e^i(t) - \ud t \sum_{j\neq i}\Bigg[ i \Big(
 a_i(\xx)\Box_{i,j}(\xx) 
+ a_j(\xx+\ii \to \xx)\Box_{i,j}(\xx)
 - \Box_{i,j}(\xx)a_i(\xx+\jj \to \xx) \nonumber \\
&- \Box_{i,j}(\xx)a_j(\xx) 
 + a_i(\xx)\Box_{i,-j}(\xx) 
 - a_j(\xx+\ii-\jj \to \xx+\ii \to \xx)\Box_{i,-j}(\xx) \nonumber \\
& - \Box_{i,-j}(\xx) a_i(\xx-\jj \to \xx)
+\Box_{i,-j}(\xx)a_j(\xx-\jj\to \xx)
 \Big) \Bigg]_\ah 
\end{align}

The parallel transported fluctuation  from site $\xx+\ii$ to site $\xx$ is denoted by
\begin{equation}
a_j(\xx+\ii \to \xx)
 \equiv
U_i(\xx)a_j(\xx+\ii)U^\dag_i(\xx),
\end{equation}
and similarly for the fields parallel transported over two links
\begin{equation}
a_j(\xx+\ii-\jj \to \xx+\ii  \to \xx)  
\equiv U_i(\xx)a_j(\xx+\ii-\jj \to \xx+\ii)U^\dag_i(\xx).
\end{equation}
In our notation there are two identical ways to write the most complicated terms involving parallel transport over two links.

\subsubsection{Gauss' law}
The Gauss' law for the fluctuations can be derived similarly as the equation of motion for the electric field fluctuation. One introduces both electric field fluctuations and gauge field fluctuations and discards the terms which are nonlinear in fluctuations. We get
\begin{equation}
 \label{eq:gauss}
 c(\xx,t)
= \sum_i \frac{1}{a^2} \Big\{ e^i(\xx)- U_i^\dagger(\xx-\ii) e^i(\xx-\ii) U_i(\xx-\ii)\nonumber
+ i U_i^\dagger(\xx-\ii)[a_i(\xx-\ii),E^i(\xx-\ii)]U_i(\xx-\ii) \Big\} =0 .
\end{equation}

\subsubsection{Equation of motion for the fluctuation of the gauge field}
The equation of motion for the gauge field fluctuation turns out to be more problematic. If one naively discretizes the continuum equation of motion (this is easy to derive by inserting the fluctuation to equation (\ref{eq:Adot})) for the gauge field fluctuation, the Gauss' law is not conserved by the time evolution. It also turns out that the linearization is broken by a term which is of second order in $\mathrm{d}t$.

The solution to this problem is to reverse engineer the correct update for the gauge field fluctuation using the Gauss's law. For the background fields, the discretized Gauss's law is conserved separately for the timesteps of links and electric fields. Demanding the same from the Gauss's law of the fluctuation we obtain an update for the fluctuation of the gauge field (for SU(2))
\begin{equation}
a_i\left(t+\mathrm{d}t\right) =  \dfrac{-i \left[E^i , \left( \Box_{0i} e_\perp^i  \Box_{0i}^\dagger  \right) \right]}{2 \mathrm{Tr}\left(E^i E^i\right)} + \Box_{0i} a_i^\perp \Box_{0i}^\dagger + \mathrm{d}t e^{i \parallel} + a_i^\parallel\left(t\right),
\end{equation}
where the time like plaquette is denoted by $\Box_{0i}=e^{i E^i\ud t}$. The notation $\perp$ and $\parallel$ refers to the components of $a_i$ and $e^i$, which are perpendicular or parallel to the $E^i$ in color space.

\subsection{Numerical tests}
\begin{figure}
\centering
\begin{subfigure}{0.48\textwidth}
\includegraphics[width=\textwidth]{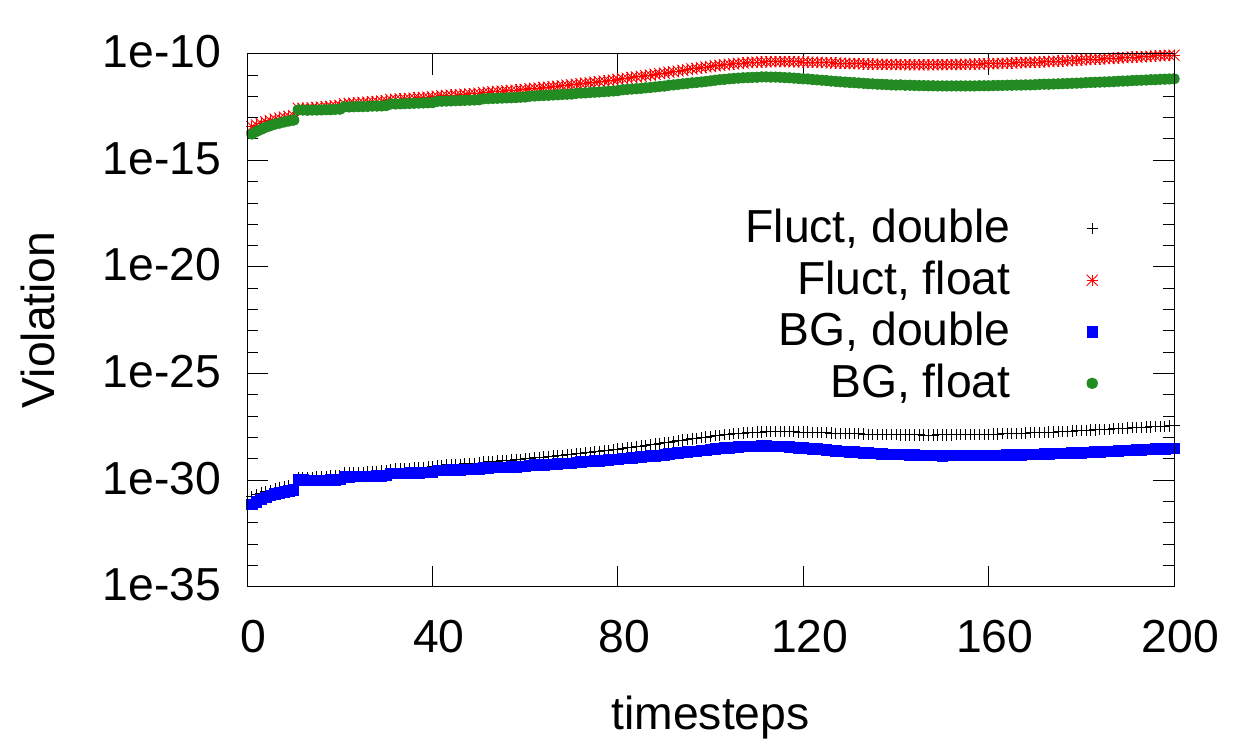}
\caption{Violation of the Gauss's law of the fluctuations as a function of time with single and double precision numbers. We have performed a random gauge transformation on every time step and fixed Coulomb gauge on every tenth timestep.}
\label{fig:Gauss}
\end{subfigure}
\hfill
\begin{subfigure}{0.48\textwidth}
\includegraphics[width=\textwidth]{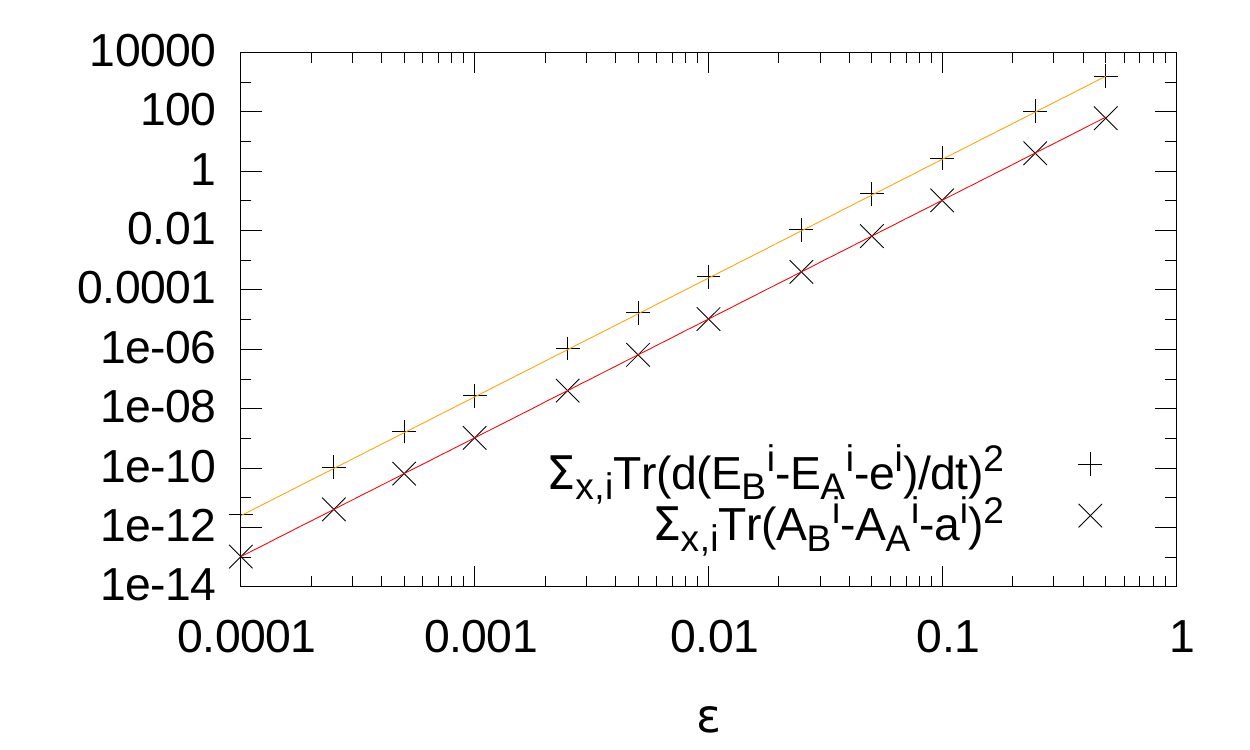}
\caption{Test of the decomposition of the field in the background field and fluctuation after some finite time. The straigth lines are fits of the form $ax^4$ demonstrating the correct power law.}
\label{fig:eps}
\end{subfigure}
\end{figure}

We have also performed a couple of simple numerical tests to verify that our equations work as they should. The main things we want to verify are that the Gauss's law is indeed conserved and that the linearization is correct. We have devised two tests. In the first one we test how well the different directions in Gauss's law cancel each other. For the background fields we use the expression:
\begin{equation} \label{eq:bggaussviol}
\frac{2\sum \limits_x \mathrm{Tr}\left(\sum \limits_{i}\left[E^i(\xx) - U^\dag_i(\xx-\ii)E^i(\xx-\ii)U_i(\xx-\ii)\right]\right)^2}{2 \sum \limits_{x,i} \mathrm{Tr}\left[E^i(\xx) - U^\dag_i(\xx-\ii)E^i(\xx-\ii)U_i(\xx-\ii)\right]^2}.
\end{equation}
And for the fluctuations we have:
\begin{equation} \label{eq:flucgaussviol}
 \frac{2 \sum \limits_x \mathrm{Tr}\left(\sum \limits_i \left[
e^i(\xx)- U_i^\dagger(\xx-\ii) e^i(\xx-\ii) U_i(\xx-\ii)
 + i U_i^\dagger(\xx-\ii)[a_i(\xx-\ii),E^i(\xx-\ii)]U_i(\xx-\ii)
\right]\right)^2 
}{
2 \sum \limits_{x,i} \mathrm{Tr} \left[
e^i(\xx)- U_i^\dagger(\xx-\ii) e^i(\xx-\ii) U_i(\xx-\ii)
 + i U_i^\dagger(\xx-\ii)[a_i(\xx-\ii),E^i(\xx-\ii)]U_i(\xx-\ii)
\right]^2}.
\end{equation} 
Here the numenator is simply the Gauss's law squared integrated over the whole lattice. The denominator is the sum of squares summed over the whole lattice. We test this using single and double precision numbers. If there is a deviation between the two, then the differences in accuracy are explained by variations in the machine precision. However, if these two will converge to same value, it means that there is a greater source of error than machine precision. Results are shown in figure \ref{fig:Gauss}. We observe that the Gauss's law is conserved much better for double precision numbers than single precision numbers, indicating that the difference is caused by finite machine precision. We also perform a random gauge transformation on every time step and fix Coulomb gauge on every tenth timestep to verify gauge invariance numerically.

The second test probes whether the linearization works correctly. We do two simulations using two different configurations. Here $E$ is the background electric field and $e$ is the fluctuation of the electric field. The scale of the fluctuation is set by a multiplicative parameter $\epsilon.$  In the first simulation we initiate the simulation by absorbing both $E$ and $e$ to the classical background and evolve this in time. The result will be labeled as $E_B$. At the initial time we have $E_B = E + e$. In another simulation we use two dynamical fields, the background field and the fluctuation field, which are evolved separately with their own solvers (The background fields appear in the equations of motion of the fluctuations of course.). We denote these fields by $E_A$ and $e_A$. Initially $E_A = E$ and $e_A = e$. After evolving these two simulations in time, we measure $\mathrm{Tr}\left(\dot{E}_B - \dot{E}_A -\dot{e}_A \right)^2,$ or alternatively $\mathrm{Tr}\left(A_B - A_A -a_A \right)^2$ for the gauge fields. If the linearization is done correctly, linear term in $\epsilon$ in the difference should cancel. Thus the traces of the squares should scale as $\epsilon^4.$ The results are shown in figure \ref{fig:eps}, where we observe the correct scaling.

\section{Conclusions}
We have studied the plasmon mass in pure glue QCD using three different methods: the effective dispersion relation, the uniform electric field method and perturbative formula relating the quasiparticle spectrum to the plasmon mass. It turns out that HTL and UE methods can be brought into rough agreement, but the DR method agrees with other methods within a factor of two. The conclusion from the agreement between the UE and HTL methods is that the kinetic theory description using weakly interacting quasiparticles seems to be valid way to understand overoccupied system of classical gauge fields, even quantitatively.

We have also developed a way to simulate dynamical fluctuations on top of classical background. We have derived, implemented and tested linearized equations for fluctuations in classical Yang-Mills theory on the lattice, and verified that they conserve the Gauss's law. 

Our future goals are to study plasmon mass in 2 dimensional systems, mimicking 2+1 dimensional boost invariant systems, which arise in the weak coupling framework in ultrarelativistic heavy ion collisions. We also have plans to apply fluctuations in our future studies. A straightforward application we are working on is to study the dispersion relation of weakly coupled plasma by using fluctuations to study linear response in CYM.

\section*{Acknowledgements}
This work has been supported by the Academy of Finland, projects 267321 and 303756, by the European Research Council, Grant ERC-2015-CoG-681707. J.P. is supported by the Jenny and Antti Wihuri Foundation. We acknowledge CSC – IT Center for Science, Finland, for computational resources.

\clearpage
\bibliography{/home/jatapeur/Dropbox/LatexBibliografia/Megabib}
\end{document}